\documentclass{elsart}

\makeatletter

\usepackage{graphicx}

\makeatother
\begin{document}

\begin{frontmatter}

\title{Factor Two Discrepancy of Hawking Radiation Temperature}

\author{Tadas K. Nakamura}

\ead{tadas@fpu.ac.jp}
\ead[url]{http://mira.bio.fpu.ac.jp/tadas/}

\address{Fukui Prefectural University, 4-1-1 Eiheiji, Fukui 910-0095, Japan}

\date{2007/10/20}

\begin{abstract}
The possibility of an alternative way to formulate the Hawking radiation
in a static Schwarzschild spacetime has been explored. To calculate
the Hawking radiation, there can be two possible choices of the spacetime
wedge pairs in the Krucal-Szekeres coordinates. One is the wedge pair
consists of exterior spacetime of a black hole and the exterior spacetime
of a white hole, and the other is that of exterior and interior spacetimes
of one black hole. The radiation from the former is the Hawking's
original one. Though the the latter has been often regarded as the
same phenomena as the former, the result here suggests it is not;
its radiation has a temperature twice as high as the Hawking temperature.
\end{abstract}


\end{frontmatter}

\maketitle

\section{Introduction }

Right after the Hawking's original paper \cite{hawking}, attempts
have been made to derive the Hawking radiation from a vacuum in
a static Schwarzschild spacetime \cite{hartle,unruh,israel,fulling,kay}.
The choice of the correct vacuum is crucial in such attempts. It is
well known that the definition of a vacuum is not unique in curved
spacetimes, and there is no general prescription to define a natural
vacuum. Therefore we have to determine somehow which vacuum is actually
realized, depending on each problem. The difficulty to find such
a vacuum in a Schwarzschild spacetime comes from its bifurcating Killing
horizons. The horizons divide the extended Schwarzschild spacetime
into four spacetime wedges, and we have several choices of wedges
or wedge pairs to define the vacuum.

Unruh \cite{unruh} compared two distinct definitions of vacua, called
$\eta$ definition and $\xi$ definition in his paper. The vacuum
in $\xi$ definition is often referred as the Hartle-Hawking vacuum,
which is calculated from analytic functions across the exterior spacetime
of a black hole and the exterior spacetime of a white hole. On the
other hand, it is also possible to define a vacuum state on the
exterior wedge of a black hole alone, which is the $\eta$ definition;
the vacuum with this definition is often called Boulware vacuum. Unruh
\cite{unruh} considered the Haretle-Hawking vacuum is preferable
because Boulware vacuum has singularity on physical quantities such
as energy. This point has been examined extensively by several authors
\cite{fulling,kay}, and it was shown that under some basic assumptions,
the Hartle-Hawking vacuum is the only mathematically reasonable one
in a wide class of spacetimes with Killing horizons \cite{kay2}.

The legitimacy of the Hartle-Hawking vacuum shown in the above mentioned
papers is, however, based on the particle number (or a conserved quantity
along the Killing field, in general) defined on a Cauchy surface in
the Kruscal-Szekeres coorinates. The vacua are calculated with analytic
functions across the two exterior spacetime wedges; one is of a black
hole and the other is of a white hole. Therefore, what have shown
is that the Hartle-Hawking vacuum is the natural vacuum among the
vacua across these two exterior spacetime wedges.

In the present paper we explore another possibility of vacuum that
spans across the interior and exterior spacetime wedges of a black
hole; we will call this R-F vacuum hereafter. This is the vacuum defined
with the particle number on a surface with $t=\textrm{constant}$.
There have been several papers \cite{unruh2,Fredenhagen,jacobson}
that regard the R-F vacuum as the source of the Hawking radiation,
however, it seems its difference from the Hartle-Hawking vacuum was
not well recognized; most of papers consider the radiation from the
R-F vacuum is the same phenomena as that from the Hartle-Hawking vacuum.
However, our calculation shows the temperature of the radiation from
the R-F vacuum is twice as large as the one with the Hartle-Hawking
vacuum, which means the R-F vacuum is not identical to the Hartle-Hawking
vacuum.

This temperature discrepancy of factor two was first reported by Akhmedov
et al. \cite{akhmedov}. In relatively recent years, an approach based
on the tunneling effect has been extensively investigated (\cite{parikh,akhmedov}
and references therein). Akhmedov et al.\ \cite{akhmedov} have carefully
examined the integration contour in the tunneling calculation, and
concluded the resulting radiation temperature is twice as large as
the Hawking's original value. The view point with tunneling effect
may not physically well founded, however, it can be reinterpreted
in the context of canonical quantization in the R-F wedge pair when
back reaction of the particles to the metric is neglected. Then what
calculated by Akhmedov et al. \cite{akhmedov} (or other papers with
tunneling picture) coincides with the radiation from the R-F vacuum
in the present paper. Therefore, the factor two discrepancy reported
by Akhmedov et al. \cite{akhmedov} can be explained by the difference
of vacua (R-F or Hartle-Hawking) from which the radiation comes from.

We first examine the case of Unruh effect in the next section. Though
the R-F vacuum has little physical significance in the Minkowski spacetime,
its radiation is mathematically simpler and can be a good example
for this type of vacua in other spacetimes. Its results are readily
applied to the Schwarzschild spacetime since the structure near the
bifurcating horizons are the same in both spacetimes. The application
to the Schwarzschild spacetime is then sketched in the following
section. A brief discussion on the validity of our approach is provided
in the last section of this paper.

\section{Radiation from Minkowski Vacua}

\subsection{Canonical Quantization with Horizons}

Suppose two dimensional coordinate system $(\eta,\xi)$ with the following
metric:\begin{equation}
ds^{2}=A(\xi)\, d\eta^{2}-B(\xi)^{-1}d\xi^{2}.\label{eq:metric}\end{equation}
 The wave equation of a massless scalar particle may be written as\begin{equation}
\frac{1}{A}\frac{\partial^{2}}{\partial\eta^{2}}\phi-\sqrt{\frac{B}{A}}\,\frac{\partial}{\partial\xi}\sqrt{AB}\,\frac{\partial}{\partial\xi}\phi=0.\label{eq:wave}\end{equation}
Now let us assume there is one and only one point where $A(\xi)B(\xi)=0$.
We choose $\xi$ coordinate such that $\xi=0$ at that point; we call
$\xi>0$ region {}``positive side'' and $\xi<0$ region {}``negative
side''.

Separating the variables, we write eigenfunctions with respect to
$\xi$ on the positive/negative side as

\begin{equation}
u_{k}^{P}=\left\{ \begin{array}{ll}
u_{k}^{P} & \;\;(\xi\ge0)\\
0 & \;\;(\xi<0)\end{array}\right.,\;\;\; u_{k}^{N}=\left\{ \begin{array}{ll}
0 & \;\;(\xi\ge0)\\
u_{k}^{N} & \;\;(\xi<0)\end{array}\right.\;.\end{equation}
 We assume the inner product $\left\langle \cdots\right\rangle $
is properly defined from the metric, and $u_{k}^{P,N}$ are normalized
as \begin{equation}
\left\langle u_{k}^{P},u_{k'}^{P}\right\rangle =\left\langle u_{k}^{N},u_{k'}^{N}\right\rangle =\delta_{kk'}\,.\end{equation}

We can construct a solution $e^{-ik\eta}U_{k}(\xi)$ that satisfies
Eq (\ref{eq:wave}) over $-\infty<\xi<\infty$ across $\xi=0$ in
the following:\begin{equation}
U_{k}=\left\{ \begin{array}{ll}
\theta_{k}^{P}\, u_{k}^{P} & \;\;(\xi\ge0)\\
\theta_{k}^{N}\, u_{k}^{N} & \;\;(\xi<0)\end{array}\right..\label{eq:uk}\end{equation}
 The wave equation Eq (\ref{eq:wave}) is satisfied when we adjust
the coefficients $\theta_{k}^{P,N}$ so that $U_{k}(\xi)$ becomes
analytic across $\xi=0$; at the same time $\theta_{k}^{P,N}$ should
satisfy the normalization condition $\left\langle U_{k},U_{k}\right\rangle =1$.
Then we can expand a wave solution $\phi$ as\[
\phi(\eta,\xi)=\sum_{k}a(\eta)\, U_{k}(\xi)\]
 Decomposing $a(\eta)$ into the positive and negative frequency modes
and calculating Boglubov coefficients from $\theta_{k}^{P,N}$, we
can obtain the spectrum of Hawking/Unruh radiation,

\subsection{R-L (Right-Left) vacuum}

This vacuum corresponds to the Hartle-Hawking vacuum in a extended
Schwarzschild spacetime; we call this type of vacuum R-L vacuum in
this paper. Obviously this vacuum is the usual vacuum realized in
a flat spacetime.

Let us define the Rindler coordinate system as \begin{equation}
t=\xi\sinh a\eta,\;\; x=\xi\cosh a\eta\,,\label{eq:rindler}\end{equation}
where $t$ and $x$ are ordinary time and space coordinates in the
Minkowski spacetime. The range of the space coordinate $\xi$ spans
$-\infty<\xi<\infty$ so that the above equation covers both R and
L regions (referred as Wedge R and Wedge L hereafter) illustrated
in Figure 1. \bigskip{}

\begin{center}
\includegraphics{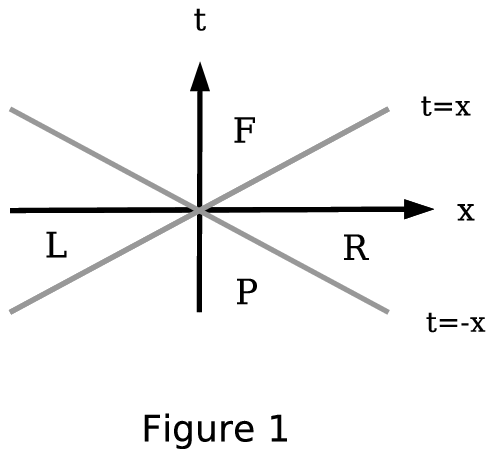} 
\par\end{center}

\begin{center}
\bigskip{}

\par\end{center}

\noindent The inner product is defined as\begin{equation}
\left\langle \phi,\phi'\right\rangle =\int_{-\infty}^{\infty}\xi^{-1}\,\phi\,\phi'^{*}\, d\xi\,.\label{eq:prod}\end{equation}
 and the eigenfunctions on Wedges R and L become\begin{equation}
u_{k}^{R}=\exp\left(\frac{ik}{a}\,\ln\xi\right),\;\; u_{k}^{L}=\exp\left(-\frac{ik}{a}\,\ln|\xi|\right)\,.\end{equation}
It should be noted that the norm of the above eigenfunctions calculated
from Eq (\ref{eq:prod}) diverges as usually we encounter in this
type of calculations. In the following we assume an appropriate prescription,
such as wave packet treatment, has been applied implicitly to avoid
this difficulty.

We wish to obtain the solution across R and L Wedges in the form of
$U_{k}$ in Eq (\ref{eq:uk}). To this end, we must choose $\theta_{k}^{P,N}$
such that Eq (\ref{eq:wave}) holds across the point of $\xi=0$.
If we choose arbitrary $\theta_{k}^{P,N}$ then discontinuity occurs
at $\xi=0$, and the right hand side of Eq (\ref{eq:wave}) will have
a $\delta$-function shaped {}``source term'' at $\xi=0$. To avoid
this we have to adjust $\theta_{k}^{P,N}$so that $U_{k}$ becomes
analytic across $\xi=0$. Using the analytic continuation of the logarithmic
function, $\ln(-\xi)=i\pi+\ln\xi$, such coefficients $\theta_{k}^{P,N}$
can be calculated in the following:\begin{equation}
\theta_{k}^{P}=\left\{ \begin{array}{ll}
{\displaystyle \frac{1}{\sqrt{1-e^{-\pi k/a}}}} & \;\;(k\ge0)\\
{\displaystyle \frac{e^{\pi k}}{\sqrt{1-e^{\pi k/a}}}} & \;\;(k<0)\end{array}\right.,\quad\theta_{k}^{N}=\left\{ \begin{array}{ll}
{\displaystyle \frac{e^{-\pi k/a}}{\sqrt{1-e^{-\pi k/a}}}} & \;\;(k\ge0)\\
{\displaystyle \frac{1}{\sqrt{1-e^{\pi k/a}}}} & \;\;(k<0)\end{array}\right.\;.\label{eq:coeffs}\end{equation}
 The Unruh radiation spectrum becomes\begin{equation}
P(k)\propto\frac{1}{2}(\theta_{k}^{P\,2}+\theta_{k}^{N\,2}-1)=\frac{1}{\exp(2\pi k/a)-1}\label{eq:unruh}\end{equation}
 for $k>0$ (see, eg., Birrel and Davies \cite{birrell}, p105 for
details of calculation).

\subsection{R-F (Right-Future) vacuum}

The above calculation has been done with the solution that spans over
Wedges of R and L in Figure 1. We perform the same calculations with
the solution over Wedges R and F in this subsection. This does not
have physical significance for a flat spacetime, however, the similar
calculation becomes important in the Schwarzschild spacetime as we
will see in the next section. We examine the R-F case with a flat
spacetime because it has essentially the same but mathematically simpler
spacetime structure.

We define the coordinates $(\eta',\xi'$) in Wedge F by\begin{equation}
t=\xi'\cosh a\eta',\;\; x=\xi'\sinh a\eta'\,,\label{eq:rindler2}\end{equation}
 where $\eta'$ and $\xi'$ are real numbers, The eigenfunction with
respect to $\xi'$ in Wedge F becomes\begin{equation}
u_{k}^{F}=\exp\left(\frac{ik}{a}\,\ln\xi'\right)\,.\end{equation}
 Let us recall that the key point in the previous calculation is in
the process to construct the solution $U_{k}$ that satisfies Eq (\ref{eq:wave})
across the singular point of $\xi=0$. There we utilized the analyticity
of $U_{k}$ as a function of $\xi$ across the both wedges. The reason
why this works is that the coordinate $\xi$ is itself analytic across
the wedges, in other words, we can express any point in both wedges
with the same single expression of Eq (\ref{eq:rindler}).

We wish to take the same approach here, that is, to express points
in Wedge F with Eq (\ref{eq:rindler2}). This can be done by complexifing
$\eta$ and $\xi$ as\begin{equation}
\eta=\frac{i\pi}{2}-\eta',\quad\xi=i\xi'\,,\end{equation}
 then the complex numbers $(\eta,\xi)$ can express any points in
Wedges F and R with Eq (\ref{eq:rindler}). Having done that, $u_{k}^{F}$
may be expressed using logarithmic analytic continuation ($\ln i\xi'=i\pi/2+\ln\xi'$)
as\begin{equation}
u_{k}^{F}=\exp\left(\frac{ik}{a}\,\ln\xi\right)=e^{\pi k/2a}\exp\left(\frac{ik}{a}\,\ln\xi'\right)\,.\end{equation}
 Then the same calculation as in the R-L case gives the radiation
spectrum as\begin{equation}
P(k)\propto\frac{1}{\exp(\pi k/a)-1}\,,\end{equation}
 which has the temperature twice as large as in the R-L case.

\section{Schwarzschild coordinates}

Now let us move on to the Schwarzschild coordinate system whose metric
is given by\begin{equation}
ds^{2}=-\left(1-\frac{2M}{r}\right)dt^{2}+\left(1-\frac{2M}{r}\right)^{-1}dr^{2}+r^{2}d\Omega^{2}\,.\end{equation}
where symbols have conventional meaning. The Rindler coordinates $(\eta,\xi)$
correspond to $(t,r-2M)$ in the above Schwarzschild coordinates.
Wedges R and F in Figure 1 correspond to the interior and exterior
spacetimes of a black hole respectively, and Wedges L and P correspond
to the {}``white hole'' in the extended Schwarzschild coordinates.
Since the spacetime structure near Killing horizons are the same in
the Rindler and Schwarzschild coordinates (see, eg., \cite{wald},
p128), the arguments in the previous subsections are valid for the
Schwarzschild coordinates with $a=1/4M$. We briefly sketch in the
following the procedure of analytic continuation that leads us to
this result.

The the solutions to the wave equation Eq (\ref{eq:wave}) have the
form of \begin{equation}
u_{k}^{R,L}\propto\exp ik[\xi+2M+2M\ln(\xi/2M)]\label{eq:uk_sch}\end{equation}
where $\xi=r-2M$. The coordinates $(t,\xi)$ naturally covers Wedge
R with $\xi>0$ and Wedge L with $\xi<0$. Using $\ln(-\xi/2M)=\ln(\xi/2M)+i\pi$,
we obtain the amplitude discontinuity as $\exp(-\pi Mk)$. This means
the radiation temperature is twice as large as the Hawking's prediction.

Continuation from Wedge R to Wedge L is not that straightforward.
We notice in the above calculation the discontinuity comes from the
analytic continuation of the logarithmic function, therefore we have
to find an appropriate way to analytically extend the logarithmic
function from Wedge R to Wedge L. To this end, we express the Schwarzschild
coordinates in Wedge L as $(t',\xi')$ and examine the analytic relation
of $\ln\xi$ in Wedge R and $\ln\xi'$ in Wedge L. The following Kruscal
Szekeres coordinates $(U,V)$ are analytical across Wedges R and L,
\begin{eqnarray}
UV & = & -2M\xi\exp\left(\frac{\xi}{2M}+1\right)\nonumber \\
|U/V| & = & \exp\left(\frac{t}{2M}\right)\,,\end{eqnarray}
 therefore, expressing $\xi$ by $U$ and $V$ gives the analyticity
across the Wedges. Near $\xi=0$ we can approximate \begin{equation}
V-U\simeq\sqrt{2M\xi}\,.\end{equation}
 at $t=0$. In Wedge R this is equivalent to\begin{equation}
\ln(V-U)=\frac{1}{2}\ln2M\xi\,,\end{equation}
 since $V-U>0$; its analytic continuation to the region of $V-U<0$
(Wedge L) is\begin{equation}
\ln(V-U)=\pi i+\ln|V-U|\,.\end{equation}
 Therefore the analytic continuation of the logarithmic function from
Wedge R to Wedge L may be written as\begin{equation}
\ln\xi'=2\pi i+\ln\xi\,.\end{equation}
 The above expression inserted into Eq (\ref{eq:uk_sch}) results
in the amplitude jump of $\exp(-2\pi Mk)$. The rest of the calculation
is the same as in the Rindler case, and we obtain the temperature
predicted in the Hawking's original paper \cite{hawking}.

\section{Concluding Remarks}

The present paper has explored the possibility of an alternative
vacuum to obtain Hawking radiation; the difference is in the choice of
the wedge pair to define the vacuum. Most of past papers with a static
Schwarzschild spacetime \cite{unruh,hartle,israel,fulling,kay,kay2}
were based on the vacuum across the exterior spacetime of a black hole
and the exterior space of a white hole, which is referred as R-L
vacuum in the present paper. On the other hand, several attempts have
made to calculate the radiation with the vacuum across the interior
and exterior spacetimes of a black hole (R-F vacuum)
\cite{unruh2,Fredenhagen,jacobson,parikh,akhmedov}. To the author's
knowledge, the difference of these two vacua is not well recognized so
far, and the radiation form R-L and R-F vacua are regarded as the same
phenomena. The result of the present paper suggests these two are
distinct; the temperature of the radiation from the R-F vacuum is
twice as large as that of R-L vacuum.

At the present we do not know which (R-F or R-L) vacuum should be
realized around a black hole. The R-L vacuum requires the Kruscal
extension, or a {}``white hole'', which is usually considered unphysical.
On the other hand, the quantum construction for R-F vacuum may be
questionable. In general, the procedure of quantization in a curved
spacetime is based on a Cauchy surface with timelike normal vectors.
However, if we choose the surface of $t=\textnormal{constant}$ in
the Schwarzschild coordinates for a R-F vacuum, it is not a Cauchy
surface because its normal vectors become spacelike in Wedge F. 

There may be two approaches to avoid this problem of R-F vacuum. One
is to extend the method of quantization to be able to utilize surfaces
with spacelike normal vectors instead of Cauchy surfaces. Recently
several papers have been published proposing a quantization method
in which an arbitrary closed surface can play the role of the Cauchy
surface \cite{oeckl,conrady,doplicher}. This method may be applicable
to the quantization for R-F vacuum. The other way is to stay in the
conventional quantization with a Cauchy surface, but define the particle
number on the surface with $t=\textnormal{constant}$. Then the surface
to define the particle number can have spacelike normal vector within
the conventional framework. Now the author of the present paper is
working on this direction and the result will be reported in a forthcoming
paper.

Before closing this paper, let us briefly take a look at the original
derivation by Hawking \cite{hawking}. The R-F vacuum means the ground
state with respect to the total {}``energy'' in the interior and
exterior spacetimes. (Here {}``energy'' means the conserved quantity
that agrees with the usual energy in a flat spacetime at the region
far away from the black hole.) Hawking \cite{hawking} examined the
process of a star collapse, assuming a vacuum state long before the
star collapse remains unchanged long after the black hole formation.
With this assumption, what realized at the later time is the Hartle-Hawking
vacuum. It was the ground state at the initial time, but it is not
after the black hole formation; the ground sate at a later time is
the R-F vacuum. The author of the present paper feels the state is
likely to settle down to the ground state somehow long after the black
hole formation, however, further investigation will be required to
verify this point. The scope of the present paper is just to point
out the possibility of R-F vacuum and cannot tell which vacuum is
actually realized around black holes at the present.

\bigskip{}

\noindent Acknowledgment: The author would like to thank M. Maeno
for suggestions and discussions.

\end{document}